\documentclass{article}
\usepackage{jcappub}
\usepackage{epsfig}
\usepackage{graphicx}
\usepackage[latin1]{inputenc}
\usepackage{amsmath}
\usepackage{url}
\def\simlt{\ \raise -2.truept\hbox{\rlap{\hbox{$\sim$}}\raise5.truept   %
\hbox{$<$}\ }}
\def\simgt{\ \raise -2.truept\hbox{\rlap{\hbox{$\sim$}}\raise5.truept   %
\hbox{$>$}\ }}                                                          %

\def\be{\begin{equation}}
\def\ee{\end{equation}}
\def\newline{\hfil\break}

\def\la{\mathrel{\hbox{\rlap{\hbox{\lower4pt\hbox{$\sim$}}}\hbox{$<$}}}}
\def\ga{\mathrel{\hbox{\rlap{\hbox{\lower4pt\hbox{$\sim$}}}\hbox{$>$}}}}

\newcommand{\pd}[3]{\frac{\partial^{#3} #1}{\partial {#2}^{#3}}} 
\newcommand{\td}[3]{\frac{d^{#3} #1}{d {#2}^{#3}}} 
\renewcommand{\v}[1]{\ensuremath{\mathbf{#1}}} 
\newcommand{\gv}[1]{\ensuremath{\mbox{\boldmath $ #1 $}}} 
\renewcommand{\bar}[1]{\ensuremath{\overline{#1}}}

\title{The role of Dark Matter sub-halos in the non-thermal emission of galaxy clusters}

\author[a,1]{Paolo Marchegiani,\note{Corresponding author.}}
\author[a]{Sergio Colafrancesco}

\affiliation[a]{School of Physics, University of the Witwatersrand, Private Bag 3, WITS-2050, Johannesburg, South Africa}

\emailAdd{Paolo.Marchegiani@wits.ac.za}
\emailAdd{Sergio.Colafrancesco@wits.ac.za}

\abstract{
Annihilation of Dark Matter (DM) particles has been recognized as one of the possible mechanisms for the production of non-thermal particles and radiation in galaxy clusters. Previous studies have shown that, while DM models can reproduce the spectral properties of the radio halo in the Coma cluster, they fail in reproducing the shape of the radio halo surface brightness because they produce a shape that is too concentrated towards the center of the cluster with respect to the observed one. However, in previous studies the DM distribution was modeled as a single spherically symmetric halo, while the DM distribution in Coma is found to have a complex and elongated shape. In this work we calculate a range of non-thermal emissions in the Coma cluster by using the observed distribution of DM sub-halos. We find that, by including the observed sub-halos in the DM model, we obtain a radio surface brightness with a shape similar to the observed one, and that the sub-halos boost the radio emission by a factor between 5 and 20\%, thus allowing to reduce the gap between the annihilation cross section required to reproduce the radio halo flux and the upper limits derived from other observations, and that this gap can be explained by realistic values of the boosting factor due to smaller substructures. Models with neutralino mass of 9 GeV and composition $\tau^+ \tau^-$, and mass of 43 GeV and composition $b \bar b$ can fit the radio halo spectrum using the observed properties of the magnetic field in Coma, and do not predict a gamma-ray emission in excess compared to the recent Fermi-LAT upper limits. These findings make these DM models viable candidate to explain the origin of radio halos in galaxy clusters, avoiding the problems connected to the excessive gamma-ray emission expected from proton acceleration in most of the currently proposed models, where the acceleration of particles is directly or indirectly connected to events related to clusters merging. Therefore, DM models deserve to be better studied both from the theoretical and observational sides; the best spectral bands where it is possible to obtain better information are the radio and the gamma-ray bands, while we do not expect a strong emission in the X-ray band. 
}

\begin{document}
\maketitle

\section{Introduction}

The annihilation of Dark Matter (DM) particles has been recognized as one of the possible mechanisms of production of non-thermal particles and the relative non-thermal emission (from radio to gamma-rays) in galaxy clusters \cite{ColafrancescoMele2001,Colafrancesco2006}. Previous studies of the synchrotron radio emission originating from DM-produced electrons in galaxy clusters \cite{ColafrancescoMele2001,Colafrancesco2006,Colafrancescoetal2011,Colafrancesco2015} have shown that these models can naturally reproduce the observed steepening of the integrated flux spectrum of the Coma cluster at high frequencies ($\nu \simgt 1.4$ GHz) for realistic values of the magnetic field (i.e., $B\sim0.1-5$ $\mu$G) in the case of light neutralino DM particles (with mass $M_\chi \sim 10-40$ GeV), but they cannot reproduce the observed shape of the radio surface brightness, unless one assumes a magnetic field increasing with radius, that is however excluded by recent observations \cite{Bonafede2010}.\\
Another argument against the simple DM interpretation of the origin of non-thermal electrons in galaxy clusters is given by the recent limits on the DM annihilation cross section  obtained with Fermi-LAT through measures in galaxy clusters \cite{Ackermann2014}, in dwarf galaxies \cite{Ackermann2015}, and in the Galactic center \cite{Calore2015}, as well as with Planck from the CMB analysis \cite{Ade2015}; these results point, in fact, to values of the annihilation cross section upper limits of the order of $\langle \sigma v \rangle \sim 10^{-27}-10^{-26}$ cm$^3$ s$^{-1}$, whereas the values required to fit the radio halo flux are of the order of $10^{-25}$ cm$^3$ s$^{-1}$ or higher (see, e.g., \cite{Colafrancescoetal2011}). 
This gap is smaller if the results of another interpretation of the gamma-ray excess measured by Fermi-LAT in the Galactic center \cite{Abaza2016} are considered: in this last work, best fit values of the annihilation cross section of $2.2\times10^{-26}$ cm$^3$ s$^{-1}$ and $7.4\times10^{-26}$ cm$^3$ s$^{-1}$ for neutralino mass of 9 GeV (with $\tau^+ \tau^-$ composition) and 43 GeV (with $b \bar b$ composition)  have been found, respectively. These values of the annihilation cross section become then interesting because, in principle, they could allow a detection of DM annihilation signals for realistic values of the boosting factors due to DM substructure distribution (see discussion in \cite{Colafrancescoetal2011}).\\
Moreover, in all the previous works the DM spatial distribution was modeled as a single, spherically symmetric halo, whereas the DM distribution observed in the Coma cluster seems to have an extended elongated shape and is structured in several sub-halos \cite{Gavazzi2009,Okabe2014}.
Interestingly, Brown \& Rudnick \cite{Brown2011} noted also that the shape of the Coma radio halo morphology is more similar to the distribution of the DM rather than to the X-ray brightness map. 
It appears therefore reasonable to think that this complex and extended distribution of DM can have an impact on both the total flux and on the surface brightness distribution of the produced DM signals.

Motivated by these observations, we study in this paper  the effect of the realistic distribution of the DM sub-halos in the Coma cluster on its emission in the radio, X-ray and gamma-ray bands. Using the models for the neutralino mass and composition corresponding to the best fit cases considered by Abazajian \& Keeley \cite{Abaza2016}, we describe the spatial distribution of the DM sub-halos with the properties found by Okabe et al. \cite{Okabe2014}, and study how the distribution of these sub-halos affects the total flux and the surface brightness distribution of the non-thermal radiation. We compare these results with the observed flux spectrum and surface brightness map in the radio band, with the Fermi-LAT upper limits in the gamma-ray band, and we finally study the possibility to detect this emission in the X-ray band.

The structure of the paper is the following: we discuss in Sect. 2 the distribution model of the DM sub-halos in Coma and we present the results of their contribution to the non-thermal radio, X-ray and gamma-ray emission in Sect.3. We discuss these results and delineate the conclusions of our study in the final Sect.4.
Throughout the paper, we use a flat, vacuum--dominated cosmological model following the results of Planck, with $\Omega_m = 0.308$, $\Omega_{\Lambda} = 0.692$ and $H_0 =67.8$ km s$^{-1}$ Mpc$^{-1}$ \cite{Ade2015}. With these values the luminosity distance of the Coma cluster at $z=0.023$ is  $D_L=104$ Mpc, and 1 arcmin corresponds to 28.9 kpc.

\section{The DM distribution model in Coma}

\subsection{DM density properties of the sub-halos}

We consider here all the DM sub-halos found by Okabe et al. \cite{Okabe2014} in the region of the central part of the radio halo in Coma, i.e. all the sub-halos at a distance less than 30 arcmin from the center of the cluster: specifically, we consider the DM sub-halos labelled with numbers 11, 12, 14, 15, 17, 18, 19, 20, 21, 23, 24, 27, 28, 29, and 31 that are listed in Table 3 of Okabe et al.\cite{Okabe2014}, where the position and the mass of each sub-halo are given. The main DM halo of the Coma cluster is centered on the sub-halo 21, and has a mass of $1.24\times10^{15}$ M$_\odot$ (see, e.g., \cite{Okabe2014}).

Following Okabe et al. \cite{Okabe2014}, we model the spatial distribution of the DM in each sub-halo with a truncated Navarro-Frenk-White profile (TNFW), i.e. a DM density distribution that follows a NFW profile until a truncation radius $r_t$ set by the effect of tidal interactions between the sub-halo and the main cluster halo, and is zero outside $r_t$:
\begin{equation}
\label{eq.tnfw}
\rho_{TNFW}(r)= \left\{ \begin{array}{ll}
\frac{\rho_s}{(r/r_s)(1+r/r_s)^{2}} & r\leq r_t \\
 0 & r>r_t \end{array} \right. \; .
\end{equation}
The scale radius  $r_s$ of the DM density profile and the halo characteristic density $\rho_s$ for each sub-halo are determined from the sub-halo mass $M_{sub}$ following the procedure described in Bullock et al. \cite{Bullock2001} and Colafrancesco, Marchegiani \& Beck \cite{Colafrancesco2015}, where the concentration parameter of an halo, defined as $c_{vir}=R_{vir}/r_s$, where $R_{vir}$ is the virial radius of the sub-halo, can be derived from a fit to the results of cosmological simulations \cite{Bullock2001,Coe2010}:
\begin{equation}
c_{vir}=\frac{9}{1+z}\left(\frac{M_{sub}}{1.3\times10^{13}h^{-1}M_\odot}\right)^{-0.13} \;.
\end{equation}
The virial radius of the sub-halo is given by:
\begin{equation}
R_{vir}^3=\frac{M_{sub}}{\frac{4}{3}\pi\Delta_c\rho_{crit}} \;,
\end{equation}
where we assume $\Delta_c=100$, and where $\rho_{crit}=2.7755\times10^2 h^2 M_\odot$ kpc$^{-3}$ is the critical density of the universe.
The characteristic density $\rho_s$ is obtained from the relation:
\begin{equation}
\frac{\rho_s}{\rho_{crit}}=\frac{\Delta_c}{3}\frac{c_{vir}^3}{\ln(1+c_{vir})-\frac{c_{vir}}
{1+c_{vir}}} \;.
\end{equation}
The truncation radius $r_t$ in eq.(\ref{eq.tnfw}) has been found by Okabe et al. \cite{Okabe2014} to be related with the sub-halo mass according to the relation $M_{sub} \propto r_t^{\alpha}$; in this respect, we used the best fit value they found for the TNFW profile, $\alpha=1.18$, to estimate the truncation radius of each sub-halo, and we normalized this relation to the values of $M_{sub}$ and $r_t$ to the best fit values reported in the second row of their Table 4, because most of the halos we consider have masses close to the intermediate range $(5-8)\times10^{12}$ M$_\odot$.

\subsection{Electrons equilibrium spectrum}

Given the radial profile of the DM density, and once we assume that DM particles are neutralino with mass $M_\chi$, 
the production rate of electrons and gamma-rays by DM annihilation is given by \cite{Colafrancesco2006}:
\begin{equation}
Q_i (E,r) = \langle \sigma v\rangle \sum\limits_{f}^{} \td{N^f_i}{E}{} B_f \mathcal{N}_{\chi} (r) \; ,
\label{source.term}
\end{equation}
where $i$ is the index referring to the output product (i.e. electrons/positrons or photons), $\langle \sigma v\rangle$ is the thermally-averaged neutralino annihilation cross-section, the index $f$ labels kinematically justified annihilation final states with branching ratios $B_f$ (in the following we will consider single final states depending on neutralino composition), and $\mathcal{N}_{\chi} (r) = (\rho_{TNFW} (r) )^2/(2 M_{\chi}^2)$ is the neutralino pair density. The production spectra $(dN^f_i)/(dE)$ can be calculated using the DarkSusy package \cite{Gondolo2004}.
These production rates can be further increased by the effect of smaller substructures, as we will discuss in Sect.2.3 below. 

The non-thermal electrons produced in the DM annihilation processes are subject to the effect of energy losses and spatial diffusion in the magnetized plasma of the cluster, according to the diffusion equation, that in the general form is written as (see, e.g., \cite{Colafrancesco2006} for details):
\begin{equation}
\pd{}{t}{}\td{n_e}{E}{} =  \; \gv{\nabla} \left( D(E,\v{r})\gv{\nabla}\td{n_e}{E}{}\right) + \pd{}{E}{}\left( b(E,\v{r}) \td{n_e}{E}{}\right) + Q_e(E,\v{r}) \; ,
\end{equation}
where $(dn_e)/(dE)$ is the electron spectrum, $D(E,\v{r})$ is the spatial diffusion function, $b(E,\v{r})$ is the energy-loss function and $Q_e(E,\v{r})$ is the electron source function.
The equilibrium solution in the case of spherical symmetry and assuming the energy-loss and diffusion functions have no spatial dependence is of the form
\begin{equation}
\td{n_e}{E}{} (E,r) = \frac{1}{b(E)}  \int_E^{M_\chi} d E^{\prime} \, G(E,E^{\prime},r) Q_e (E^{\prime},r) \; ,
\end{equation}
where $G(E,E^{\prime},r)$ is a Green's function, that can be neglected in halos of cluster size, but cannot be neglected in halos of smaller size (see, e.g.,  \cite{Colafrancesco2006}).
We calculate the equilibrium spectrum of these electrons by considering Coulomb, bremsstrahlung, synchrotron and inverse Compton scattering energy losses, and a diffusion coefficient appropriate for a Kolmogorov magnetic field fluctuation spectrum $D(E)=D_0 d_b^{2/3} B^{-1/3} E^{1/3}$ where $d_b\sim20$ kpc is the minimum scale of homogeneity of the magnetic field in Coma, $B$ is the magnetic field in $\mu G$, $E$ is the electrons energy in GeV and we assume $D_0=3.1\times10^{28}$ cm$^{2}$ s$^{-1}$.

As for the Coma magnetic field radial profile, we assume the best fit result found from Faraday Rotation measures studies in the Coma cluster, i.e. $B(r)=B_0 (n_{th}(r)/n_{th}(0))^\delta$ with $\delta=0.5$ and $B_0=4.7$ $\mu$G \cite{Bonafede2010}. 
As for the thermal gas density profile $n_{th}(r)$ we assume a beta-model with the parameters found by Briel, Henry \& Boehringer \cite{Briel1992}. 
Since the size of the DM sub-halos is generally small compared to the dimension of the cluster, we assume that the magnetic field and the gas density are constant within each sub-halo with the values calculated at the center of each sub-halo. With this assumption the Green's function has an analytical expression (see eq. A.13 in \cite{Colafrancesco2006}). The equilibrium spectrum in the main halo is instead calculated by neglecting the diffusion term, and using the described radial dependences for the magnetic field and the gas density.

Once the electron equilibrium spectrum in each sub-halo is derived in this way, we calculate the radio surface brightness produced by synchrotron emission in each DM sub-halo, and we center this emission on the position of the corresponding sub-halo in the map. We calculate also the total radio flux given by the sum of all the sub-halo produced radio emission. Finally, we calculate the total gamma-ray flux, as well as the flux and the surface brightness of the inverse Compton scattering emission observable in the X-ray band.

\subsection{Effect of smaller sub-halos}

The weak lensing observations used in this paper to describe the properties of DM sub-halos can only detect sub-halos with masses greater than $\sim10^{12}$ M$_\odot$ \cite{Okabe2014}. Since DM halos form hierarchically, we expect that a large number of sub-halos with smaller mass is present inside the volume of the cluster, as confirmed by cosmological simulations (e.g. \cite{Diemand2008}). Since the neutralino pair density $N_\chi$ is proportional to the square of the DM density, we can expect that the effect of these smaller sub-halos is to boost the production of secondary electrons and gamma-rays (e.g. \cite{Colafrancesco2006,Pieri2011}).

The effect of sub-halos in a DM halo of mass $M_h$ is usually described by introducing a boosting factor ${\cal B}_{sh} (M_h)$, defined as \cite{Stringari2007}:
\begin{equation}
{\cal L} (M_h) = [1+{\cal B}_{sh}(M_h)] {\cal L}_{sm}(M_h) \;,
\end{equation} 
where $\cal L$ and ${\cal L}_{sm}$ are annihilation factors related to the structures of sub-halos, given by the integration of the square of DM density over the line of sight and the solid angle, for the total and the smooth halo, respectively. Therefore, the value of ${\cal B}_{sh}$ can be found by integrating the annihilation factors of sub-halos with mass $m$ weighted by the number of sub-halos having mass $m$ over the whole range of sub-halo masses \cite{Stringari2007}:
\begin{equation}
{\cal B}_{sh} (M_h) = \frac{1}{{\cal L}_{sm}} \int_{m_0}^{M_h} \frac{dN}{dm} {\cal L}(m) \;.
\label{eq.boost}
\end{equation}
Usually, the sub-halo mass function is assumed to have a power-law shape:
\begin{equation}
\frac{dN}{dm} \propto m^{-\mu} \;,
\label{sh_mass_funct}
\end{equation}
where the normalization factor of this relation is related to the fraction of the mass included in the sub-halos \cite{Colafrancesco2006}, and where $\mu$ is quite well constrained from cosmological simulation and is in the range $\mu \sim1.9-2$ \cite{Diemand2008,Springel2008}, in agreement with theoretical expectations based on the Press-Schechter theory for structure formation \cite{Giocoli2008}.

Unfortunately, the determination of the boost factor from eq.(\ref{eq.boost}) is not immediate. In fact, the annihilation factor ${\cal L}(m)$ for a sub-halo with mass $m$ is in turn affected by the presence of sub-halos of smaller size, and these annihilation factors depend on the sub-halo properties like radial profile, extension, and concentration, that for a sub-halo inside a larger halo do not have necessarily the same properties than for isolated halos with similar mass, because of some effects like stripping due to tidal interactions. Another quantity not well known is the value of the smallest sub-halo mass $m_0$ for which the sub-halo mass function has the power-law shape as in eq.(\ref{sh_mass_funct}): a fiducial value is given by the WIMP free streaming scale, of the order of $10^{-6}$ M$_\odot$ \cite{Green2005}, but smaller values are not excluded \cite{Bringmann2009}. Given the steep shape of the sub-halo mass function, this value can strongly impact on the value of the boost factor.

As a consequence of these uncertainties, current estimates of the correct value of the boost factor in a galaxy cluster are controversial. While in some numerical high-resolution simulation of DM distribution in galaxy clusters the boost factor can reach values of 700 or more \cite{Springel2008,Gao2012}, other studies point to lower values of the order of 30--35 \cite{Colafrancescoetal2011,Anderhalden2013,Sanchez2014}, even though this value can be increased by several factors, like a steeper cusp profile in the inner part of sub-halos, that can give boost factors up to $\sim70$ \cite{Ishiyama2014}, or the effect of tidal stripping that can increase the boost factors even by a factor 2--5 \cite{Bartels2015}. Therefore, we can conclude that standard values for the boost factor in a galaxy cluster can be considered of the order of $\sim30-35$, while an optimistic estimate of the value of the boost factor in galaxy clusters can be of the order of $\sim100$. 

Given these uncertainties, in the following we parametrize the effect of the boost factor with a term $\cal B$ independent of radius, that boosts the effect of the DM annihilation signal by multiplying the source term in eq.(\ref{source.term}). Strictly speaking, the boost factor should depend on the radius, because it depends on the spatial distribution of sub-halos inside the cluster, but since it has been found that the effect of sub-halos on the main halo is to provide a total DM radial shape following a NFW profile \cite{Pieri2011}, and since we are already using a NFW profile to describe the distribution of DM in the main halo, we consider the effect of the boost factor only on the global normalization of the DM emission, assuming that it does not change heavily the surface brightness shape. We note also that the boost factor following our definition differs slightly from the one defined in eq.(\ref{eq.boost}), i.e. ${\cal B} \equiv 1+{\cal B}_{sh}$.

\section{Results}

We assume for the DM particle a neutralino with the compositions and masses corresponding to the models that best fit the Galactic center excess according to Abazajian \& Keeley \cite{Abaza2016}: these are the case with $M_\chi=9$ GeV and composition $\tau^+ \tau^-$, and the case with $M_\chi=43$ GeV and composition $b \bar b$. 
Once the neutralino mass and composition are assumed, the only quantity that remains as a free parameter for the fit to the data is the normalization, given by the DM annihilation cross section $\langle \sigma v\rangle$, eventually multiplied by the substructure boost factor $\cal B$. In our approach, we will calculate the values of ${\cal B} \times \langle \sigma v\rangle$ necessary to reproduce the observed radio flux, and subsequently we will check, given the available upper limits on the annihilation cross section, if the resulting requirements on the boost factor value are compatible with the present theoretical estimates summarized in Sect.2.3.

\subsection{The radio flux}

\begin{figure}
\begin{center}
{
 \epsfig{file=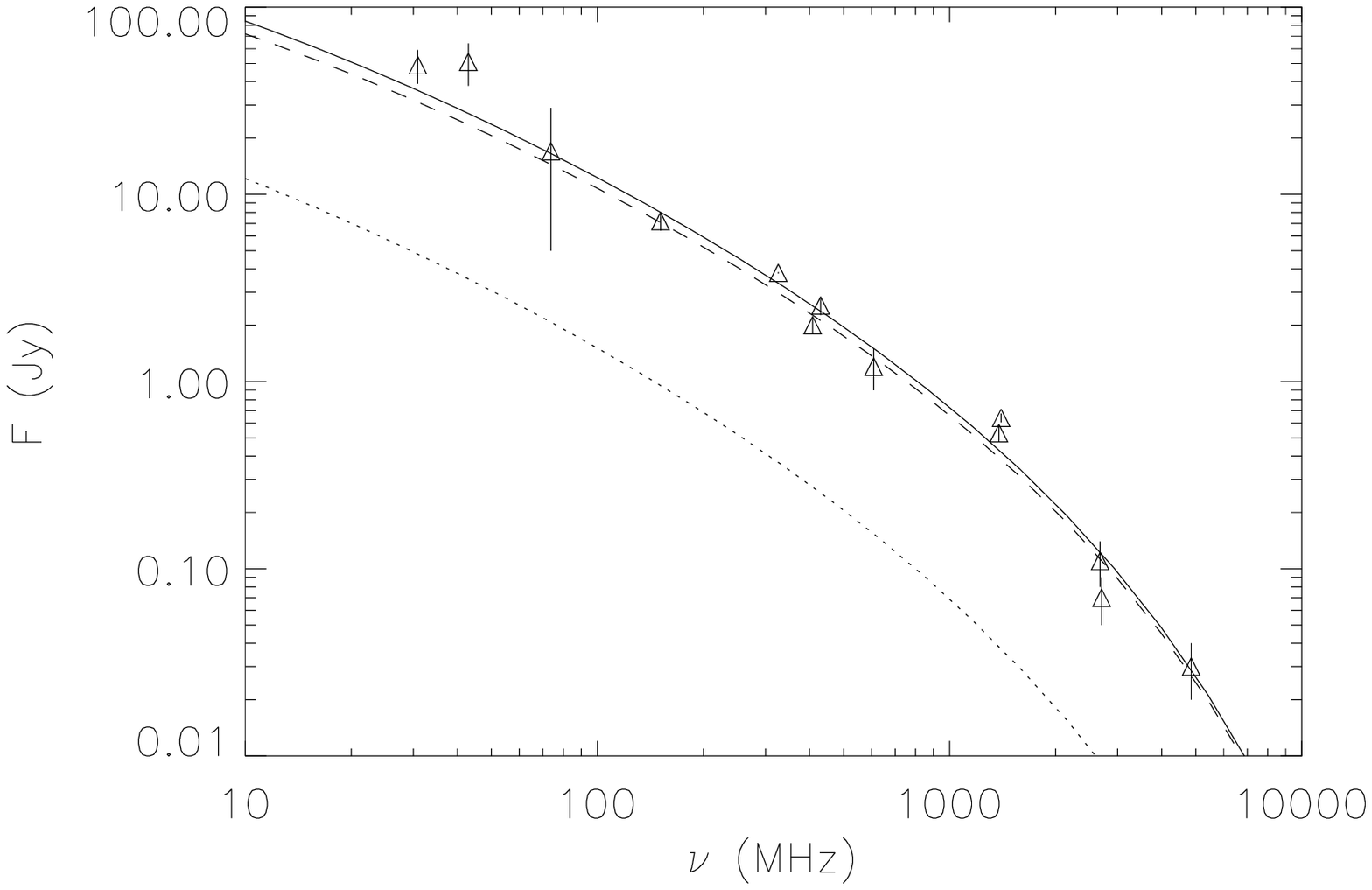,scale=0.6,angle=0.0}
 \epsfig{file=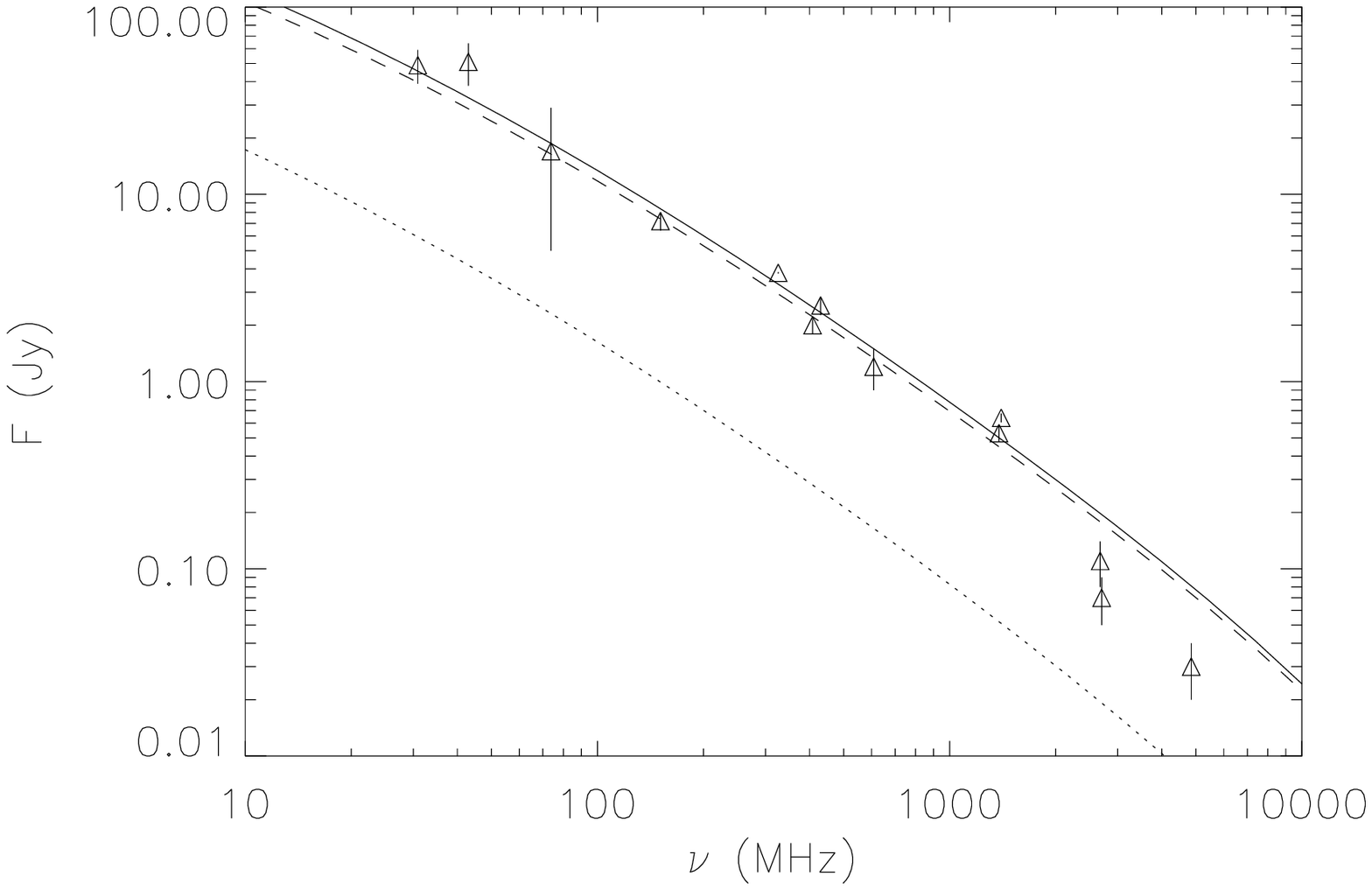,scale=0.6,angle=0.0}
}
\end{center}
 \caption{Upper panel: radio flux produced in Coma from the DM halos for a neutralino with $M_\chi=9$ GeV, composition $\tau^+ \tau^-$ and ${\cal B} \times \langle \sigma v \rangle=6\times10^{-25}$ cm$^3$ s$^{-1}$; lower panel: the same for $M_\chi=43$ GeV, composition $b \bar b$ and ${\cal B} \times \langle \sigma v \rangle=4\times10^{-24}$ cm$^3$ s$^{-1}$. In both panels with the solid line is shown the total emission, with the dashed line the contribution from the main halo and with the dotted line the contribution from the sub-halos. Data are from Thierbach et al. \cite{Thierbach2003} and references therein.}
 \label{radio_flux}
\end{figure}

For the assumed magnetic field distribution, the total spectrum of the radio halo in Coma for the two DM models we are considering is shown in Fig.\ref{radio_flux}. While the 43 GeV model is slightly better at low frequencies (i.e., around 30 MHz), the 9 GeV model reproduces better the high-frequency steepening after 1.4 GHz. In both models, most of the emission is coming from the main halo, and the sub-halos provide a contribution that is of the order of $20\%$ at small frequencies and $5-9\%$ (for the cases of 9 and 43 GeV neutralino mass, respectively) at high frequencies. For the 9 GeV neutralino case, the normalization required to reproduce the intensity of the observed flux is given by ${\cal B} \times \langle \sigma v \rangle=6\times10^{-25}$ cm$^3$ s$^{-1}$, and for the 43 GeV neutralino case the normalization is ${\cal B} \times \langle \sigma v \rangle=4\times10^{-24}$ cm$^3$ s$^{-1}$. 
In Table \ref{tab.boostfactor} we compare these values with the DM annihilation cross section upper limits found by Fermi for dwarf galaxies \cite{Ackermann2015}, Planck from CMB analysis \cite{Ade2015}, and the best fit values found by Abazajian \& Keeley \cite{Abaza2016}. Note also that previously found annihilation cross section upper limits from the Galactic center excess \cite{Calore2015} were similar to the dwarf galaxies upper limits (see discussion in \cite{Abaza2016}). 

We found that, if the DM annihilation cross section values are similar to the upper limits set by Fermi and Planck, boost factor values of the order of $\simlt 200$ and $\simgt 300$ are required for the 9 and 43 GeV neutralino models, respectively.\\
As discussed in Sect.2.3, standard values of the boost factor in galaxy clusters can be assumed to be around ${\cal B} \sim 30-35$, whereas under optimistic assumptions the boost factor can be of the order of ${\cal B}\sim100$.
We note also that a residual increase of a factor of $\sim1.5-2$ in the synchrotron emission is possible due to fluctuations in the magnetic field structure (see, e.g., \cite{Colafrancesco2005}). Therefore the values of the boost factor required to explain the observed emission for the 9 and 43 GeV neutralino models are reachable when optimistic assumptions are used, with the first one appearing less problematic than the second one. If, instead, the values of the DM annihilation cross section are those estimated by Abazajian \& Keeley \cite{Abaza2016}, the required boost factors are in the range of the expected values also in the non-optimistic cases, especially for the 9 GeV neutralino model. 

\begin{table}[htb]{}
\vspace{2cm}
\begin{center}
\begin{tabular}{|*{4}{c|}}
\hline 
Model & Source & $\langle \sigma v \rangle$ & ${\cal B}$ \\
 &  & cm$^{3}$ s$^{-1}$ &   \\
\hline 
9 GeV, $\tau^+\tau^-$ & Fermi 15 Dwarf Galaxies (UL) & $4.1\times10^{-27}$ & 150\\
 & Planck (UL) & $3.6\times10^{-27}$ & 170\\
 & Galactic center & $2.2\times10^{-26}$ & 27\\
 \hline
43 GeV, $b \bar b$ & Fermi 15 Dwarf Galaxies (UL) & $1.2\times10^{-26}$ & 330\\
 & Planck (UL) & $1.7\times10^{-26}$ & 235\\
 & Galactic center & $7.4\times10^{-26}$ & 54\\ 
 \hline
 \end{tabular}
 \end{center}
 \caption{Upper limits on the annihilation cross section from Fermi-LAT analysis of 15 dwarf galaxies \cite{Ackermann2015} and from the Planck analysis of the CMB \cite{Ade2015} and best fit values from the Galactic center excess \cite{Abaza2016}. In the last column we report the boost factor necessary to obtain the normalization factors we found, ${\cal B} \times \langle \sigma v \rangle=6\times10^{-25}$ cm$^3$ s$^{-1}$ and ${\cal B} \times \langle \sigma v \rangle=4\times10^{-24}$ cm$^3$ s$^{-1}$ for the 9 and 43 GeV neutralino models, respectively, assuming that the DM annihilation cross sections are equal to the experimental upper limits.} 
 \label{tab.boostfactor}
 \end{table}

\subsection{Gamma-ray flux}

\begin{figure}
\begin{center}
{
 \epsfig{file=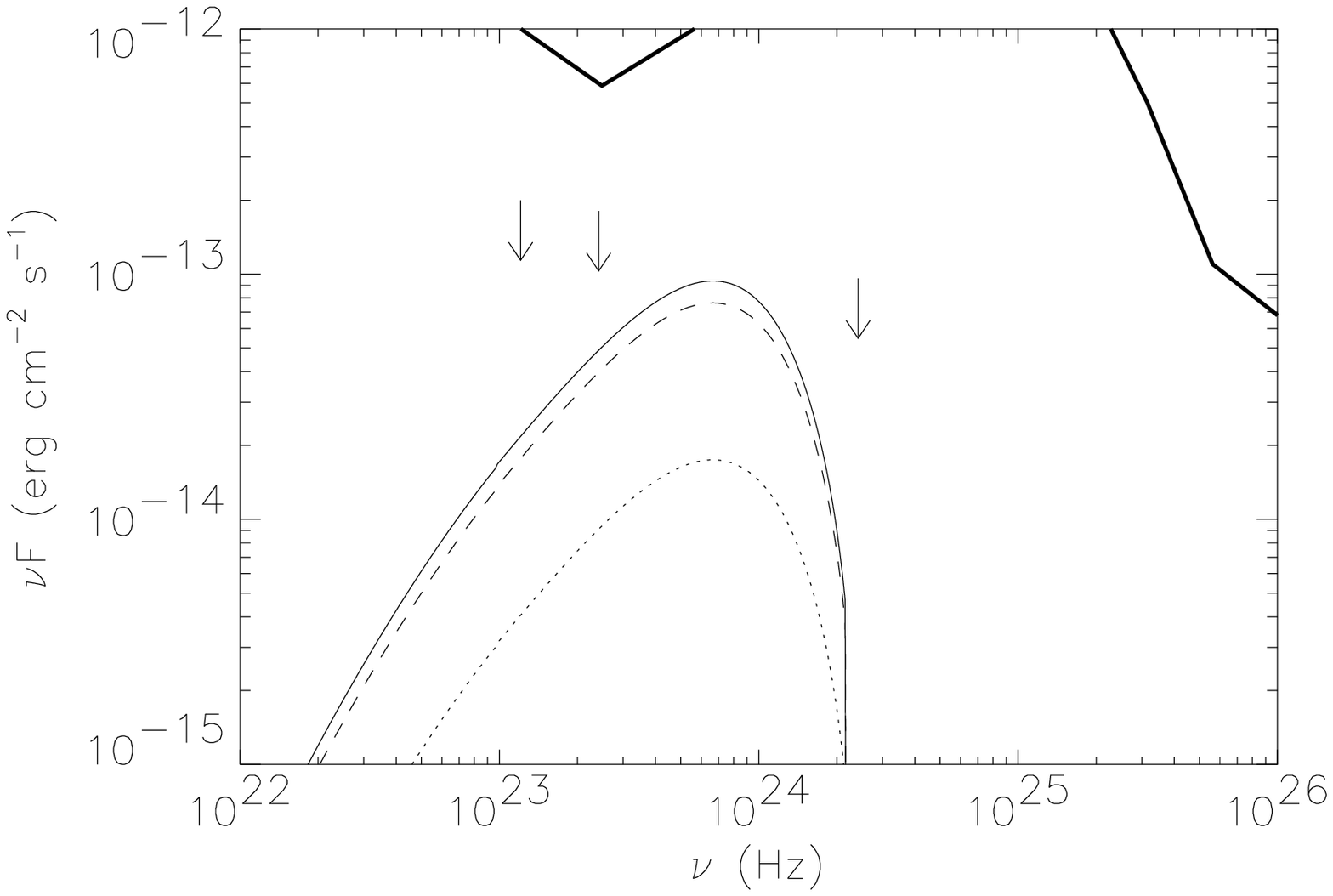,scale=0.6,angle=0.0}
 \epsfig{file=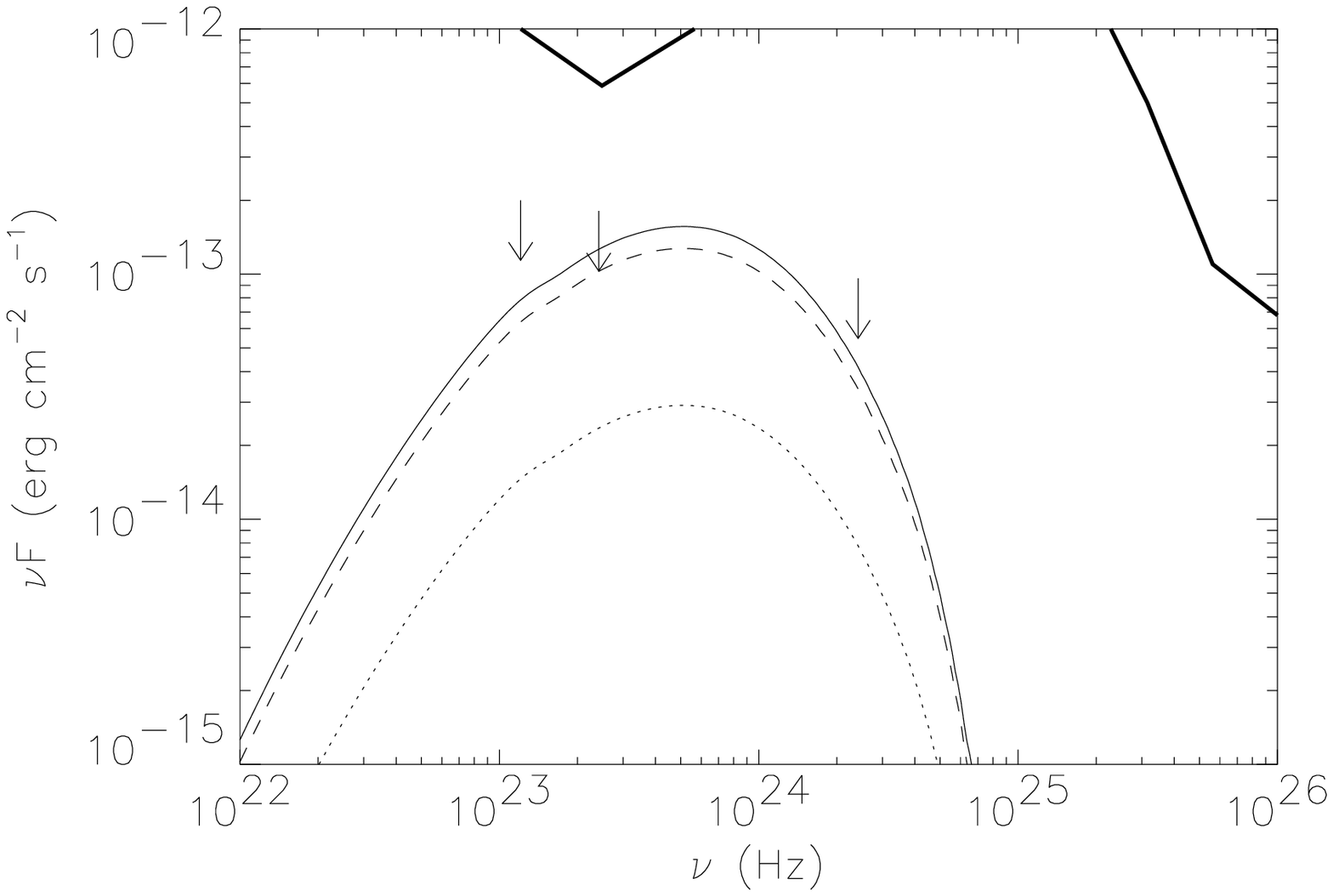,scale=0.6,angle=0.0}
}
\end{center}
 \caption{Gamma ray flux produced in Coma from the DM halos for the same models than in Fig.\ref{radio_flux}, with neutralino mass of 9 GeV (upper panel) and 43 GeV (lower panel), respectively. The solid line shows the total emission, the dashed line shows the contribution from the main halo and the dotted line shows the contribution from the DM sub-halos. Fermi-LAT upper limits are from \cite{Ackermann2014}. We also plot the expected sensitivities of ASTROGAM for an effective exposure of 1 yr (from http://astrogam.iaps.inaf.it/scientific\_instrument.html) and CTA for 1000 hrs (from \cite{Funk2013}).}
 \label{gamma_flux}
\end{figure}

When these models are used to calculate the gamma-ray emission (produced by neutral pion decay, non-thermal bremsstrahlung and inverse Compton scattering), we obtain for both models flux values which are below the upper limits set by the Fermi-LAT analysis (see Fig.\ref{gamma_flux}). In Table \ref{tab.gamma} we compare the Fermi-LAT upper limits with the calculated flux for the two DM models we are considering here.\\
We notice that the most constraining limit is the one obtained in the band $E>1$ GeV, where the gamma-ray upper limit is a factor 2.8 higher than the calculated flux for the 9 GeV neutralino model, and a factor 1.4 higher for the 43 GeV neutralino model. Fig.\ref{gamma_flux} shows that  forthcoming gamma-ray instruments  different from Fermi are not expected to have the possibility to detect this flux. At present there is a possibility to check these models by detecting this emission only if the analysis of the Fermi data from its full operative lifetime will be able to improve its upper limits by a factor 1.5 or more. 
We also note that, as discussed in the previous subsection, if there is a contribution of a factor $\sim1.5-2$ in the synchrotron emission given by the effect of magnetic field fluctuations, the corresponding gamma-ray emission must be decreased of the same factor, because the gamma-ray emission is not affected by magnetic field fluctuations. In this case, the detection of the DM-induced gamma-ray signal would be even more difficult.

\begin{table}[htb]{}
\vspace{2cm}
\begin{center}
\begin{tabular}{|*{4}{c|}}
\hline 
Band & Upper limit & Flux (9 GeV model) & Flux (43 GeV model) \\
 GeV & cm$^{-2}$ s$^{-1}$ & cm$^{-2}$ s$^{-1}$ &  cm$^{-2}$ s$^{-1}$ \\
\hline 
$>0.1$ & $4.2\times10^{-9}$ & $9.3\times10^{-11}$ & $2.7\times10^{-10}$ \\
$>0.5$ & $2.5\times10^{-10}$ & $6.1\times10^{-11}$ & $1.4\times10^{-10}$ \\
$>1$ & $1.13\times10^{-10}$ & $4.0\times10^{-11}$ & $8.0\times10^{-11}$ \\
$>10$ & $6.0\times10^{-12}$ & - & $7.4\times10^{-13}$ \\
 \hline
 \end{tabular}
 \end{center}
 \caption{Gamma upper limits set by Fermi for $E>0.1$ GeV (from \cite{Ackermann2016}) and $E>0.5$, 1, and 10 GeV (from \cite{Ackermann2014}) compared with the fluxes calculated for the 9 and 43 GeV neutralino models with the same parameters used in Fig.\ref{radio_flux}.} 
 \label{tab.gamma}
 \end{table}

\subsection{Radio surface brightness}

\begin{figure}
\begin{center}
{
 \epsfig{file=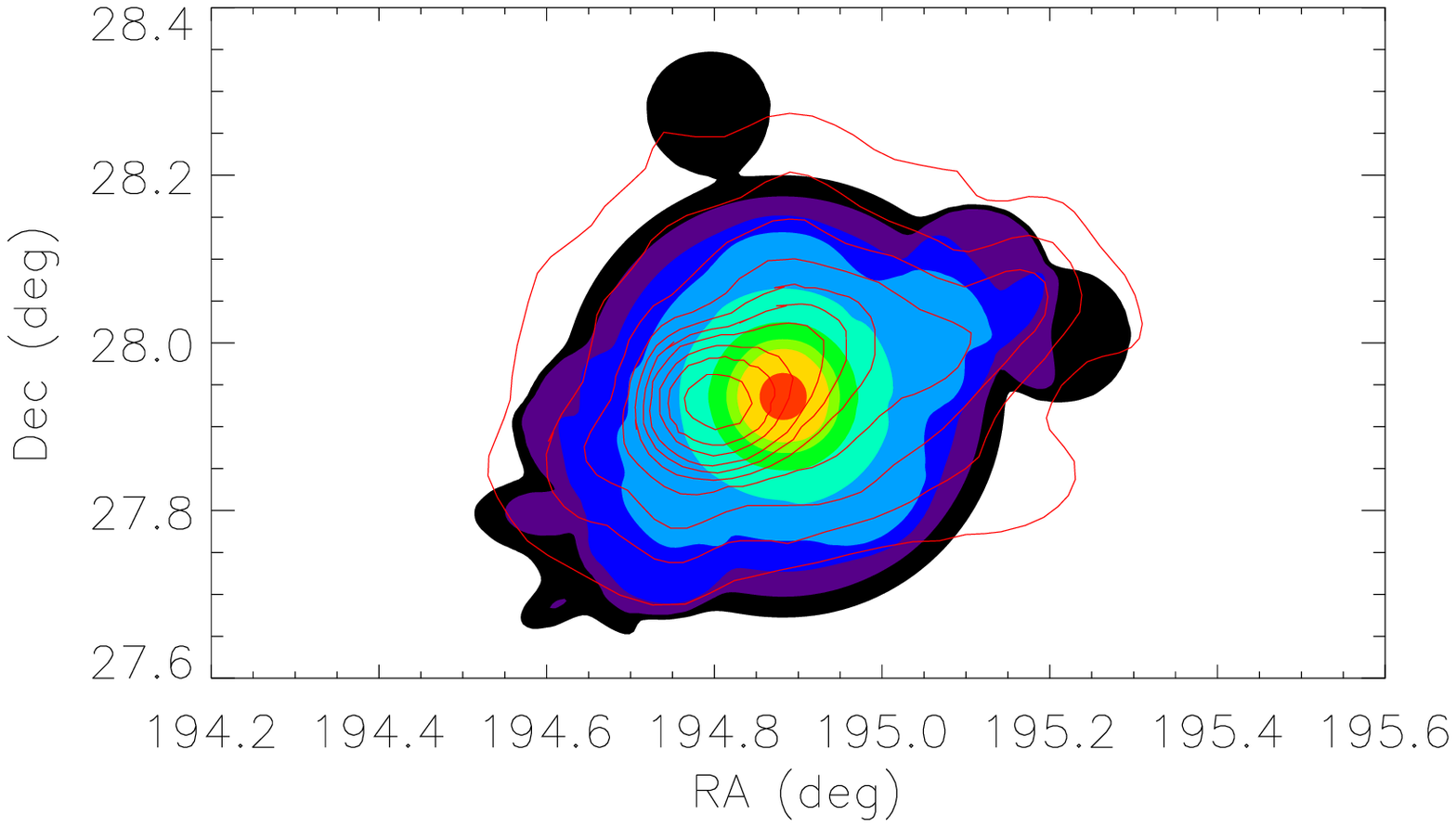,scale=0.7,angle=0.0}
 \epsfig{file=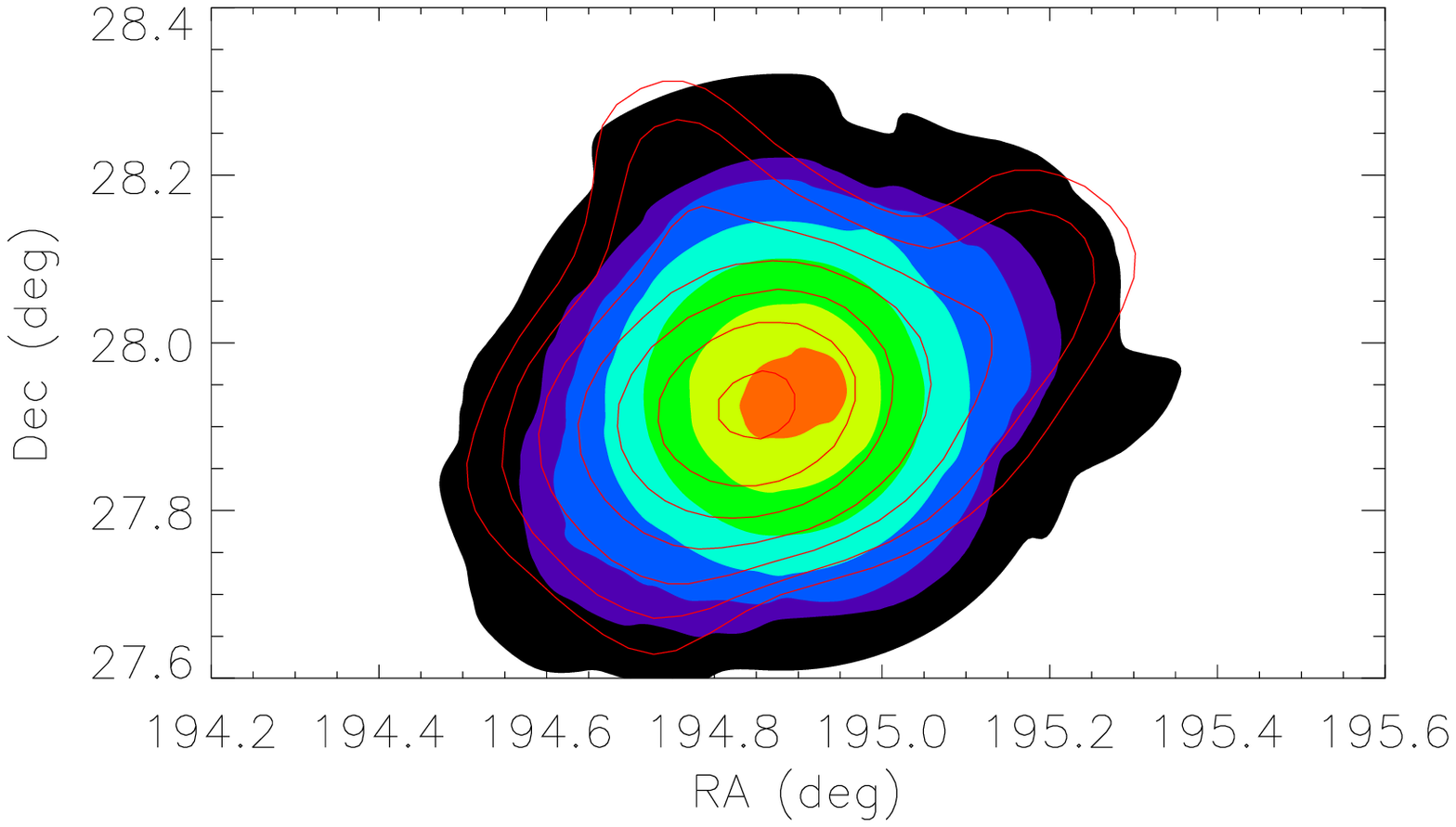,scale=0.7,angle=0.0}
}
\end{center}
 \caption{Map of the radio surface brightness in Coma from the DM halos for the 9 GeV neutralino model as in Fig.\ref{radio_flux}. Top panel: map at 352 MHz smoothed on a scale of 5 arcmin compared with the contours of Fig.10 of Brown \& Rudnick \cite{Brown2011}. Bottom panel: map at 2.675 GHz smoothed on a scale of 9.35 arcmin compared with the contours of Fig.4 of Thierbach et al. \cite{Thierbach2003}.}
 \label{radio_map}
\end{figure}

\begin{figure}
\begin{center}
{
 \epsfig{file=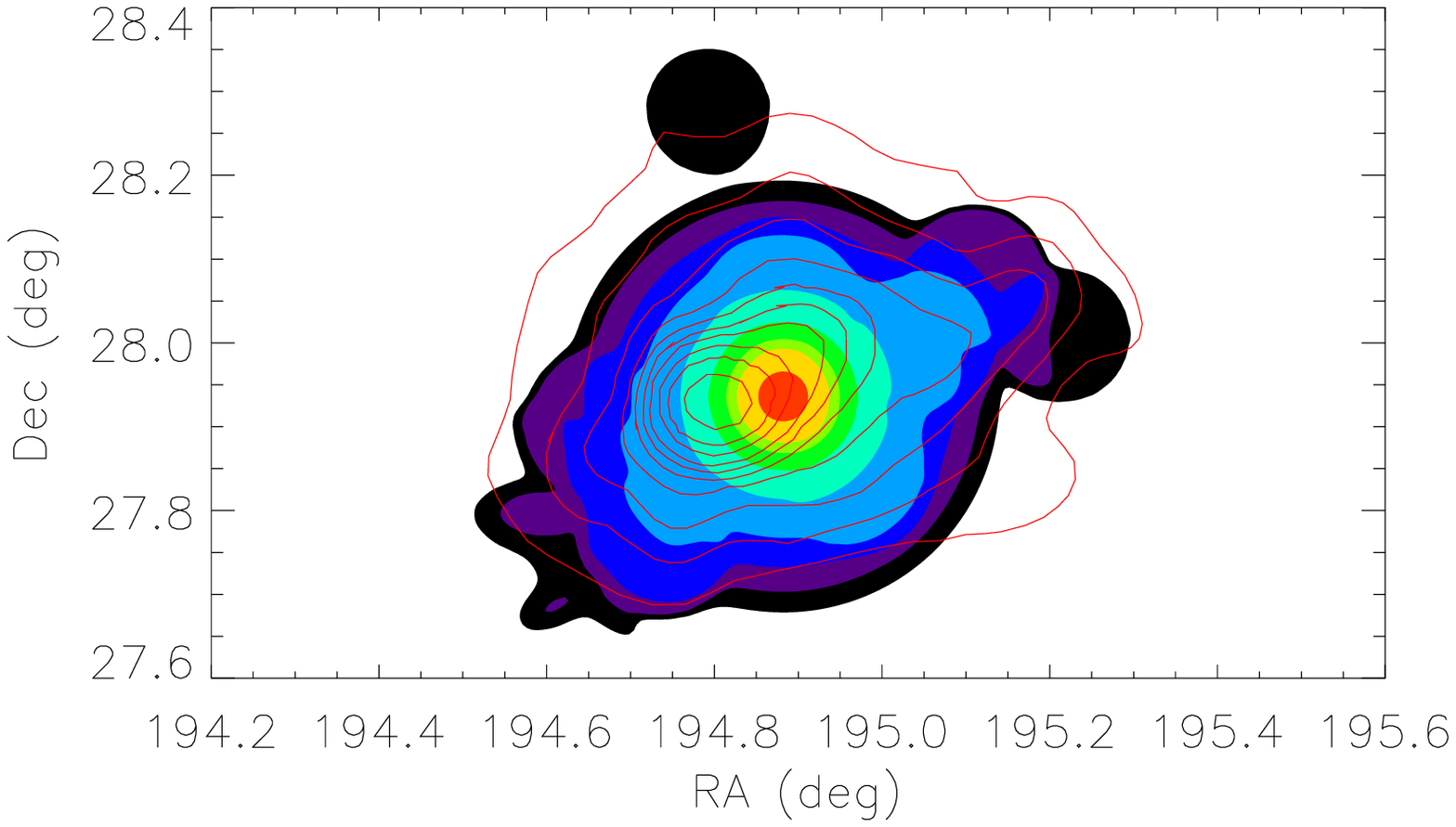,scale=0.7,angle=0.0}
 \epsfig{file=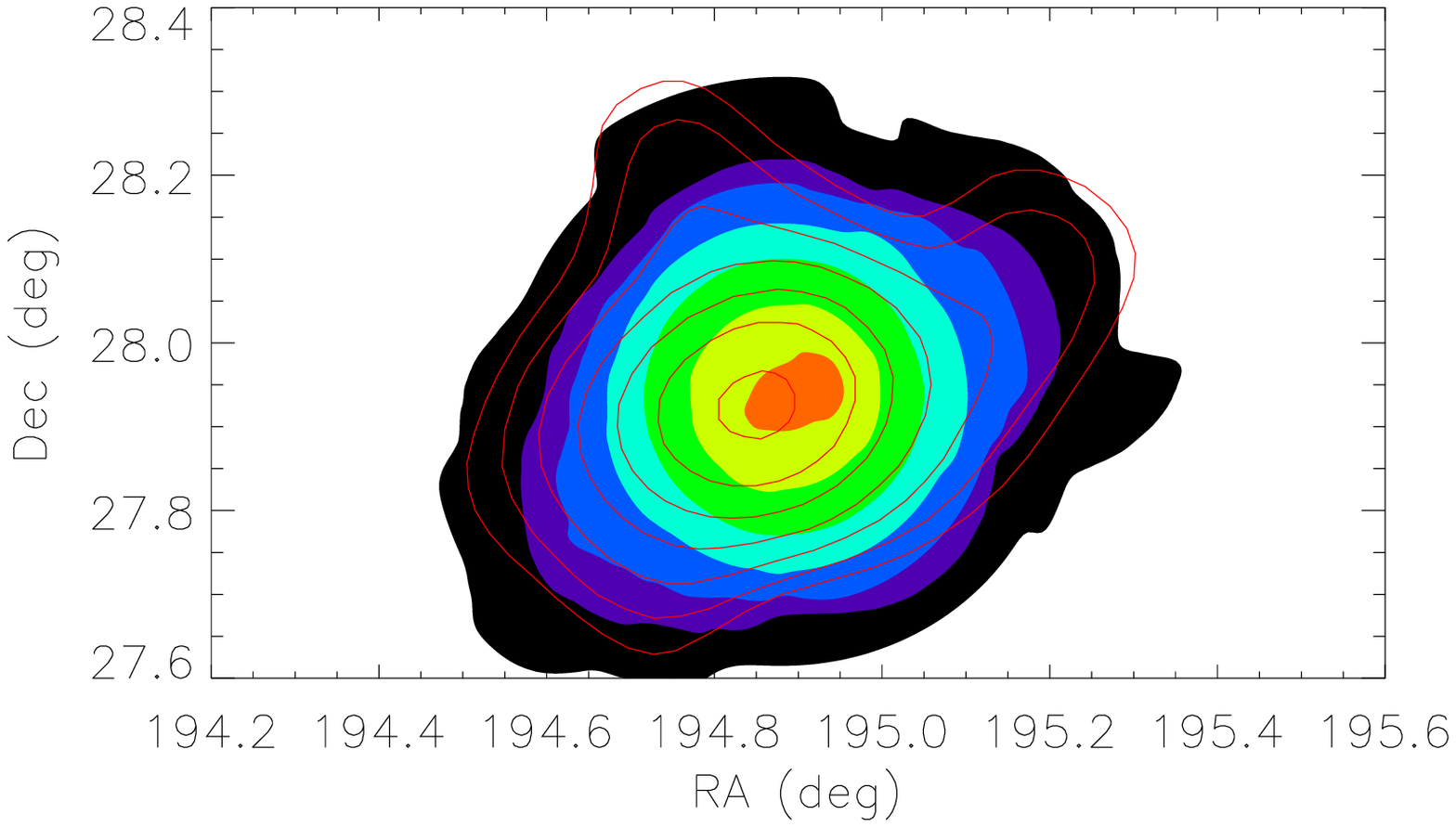,scale=0.7,angle=0.0}
}
\end{center}
 \caption{Map of the radio surface brightness in Coma from the DM halos as in  Fig.\ref{radio_map} but for the 43 GeV neutralino model.}
 \label{radio_map2}
\end{figure}

For the spatial distribution of the radio emission produced by the DM halos, we show the results for both the DM models we used. The calculated maps are shown in Figs.\ref{radio_map} and \ref{radio_map2} (the two DM models produce quite similar maps), compared with the maps of Brown \& Rudnick \cite{Brown2011} at 325 MHz, and of Thierbach, Klein \& Wielebinski \cite{Thierbach2003} at 2.675 GHz, after smoothing the map with the beam used in these papers. 
We see that the distribution of DM in Coma gives origin to a radio emission with a shape that is quite similar to the observed one (actually more similar for the 2.675 GHz map), but not exactly the same (note also that the contour levels in the calculated maps are chosen arbitrarily and do not match the levels of the observed maps). However, the procedure to obtain the maps of the diffuse radio emission is quite delicate, and depends in particular from the removal of the discrete or slightly extended sources, and can alter the real shape of the diffuse radio emission (note, for example, that the maps at low and high frequency have peaks in different positions), because by subtracting the discrete source emission it is possible that also part of the DM emission in the same location is subtracted (see, e.g., a similar discussion for the Bullet cluster in \cite{Marchegiani2015}).

\begin{figure}
\begin{center}
{
 \epsfig{file=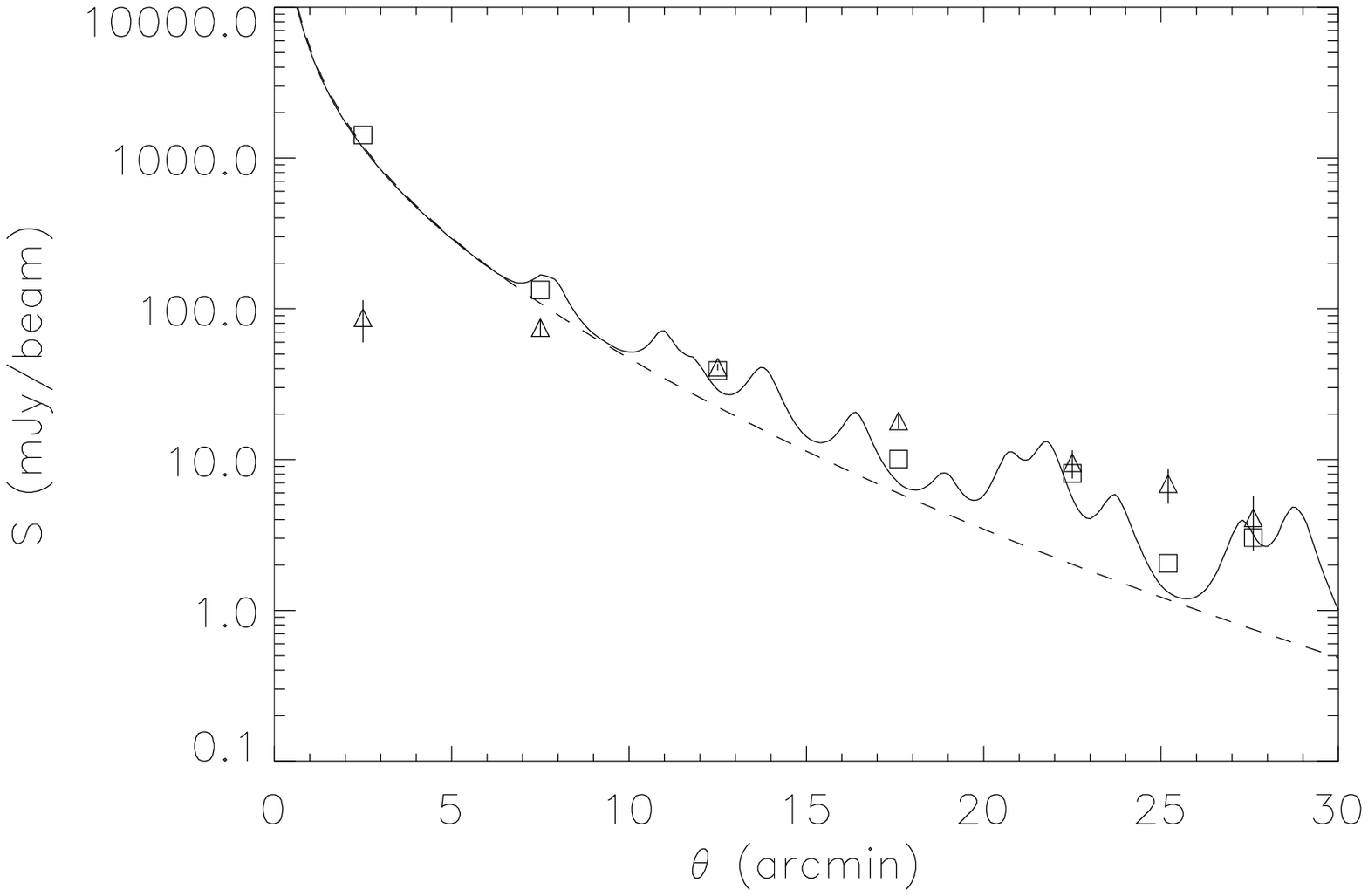,scale=0.6,angle=0.0}
 \epsfig{file=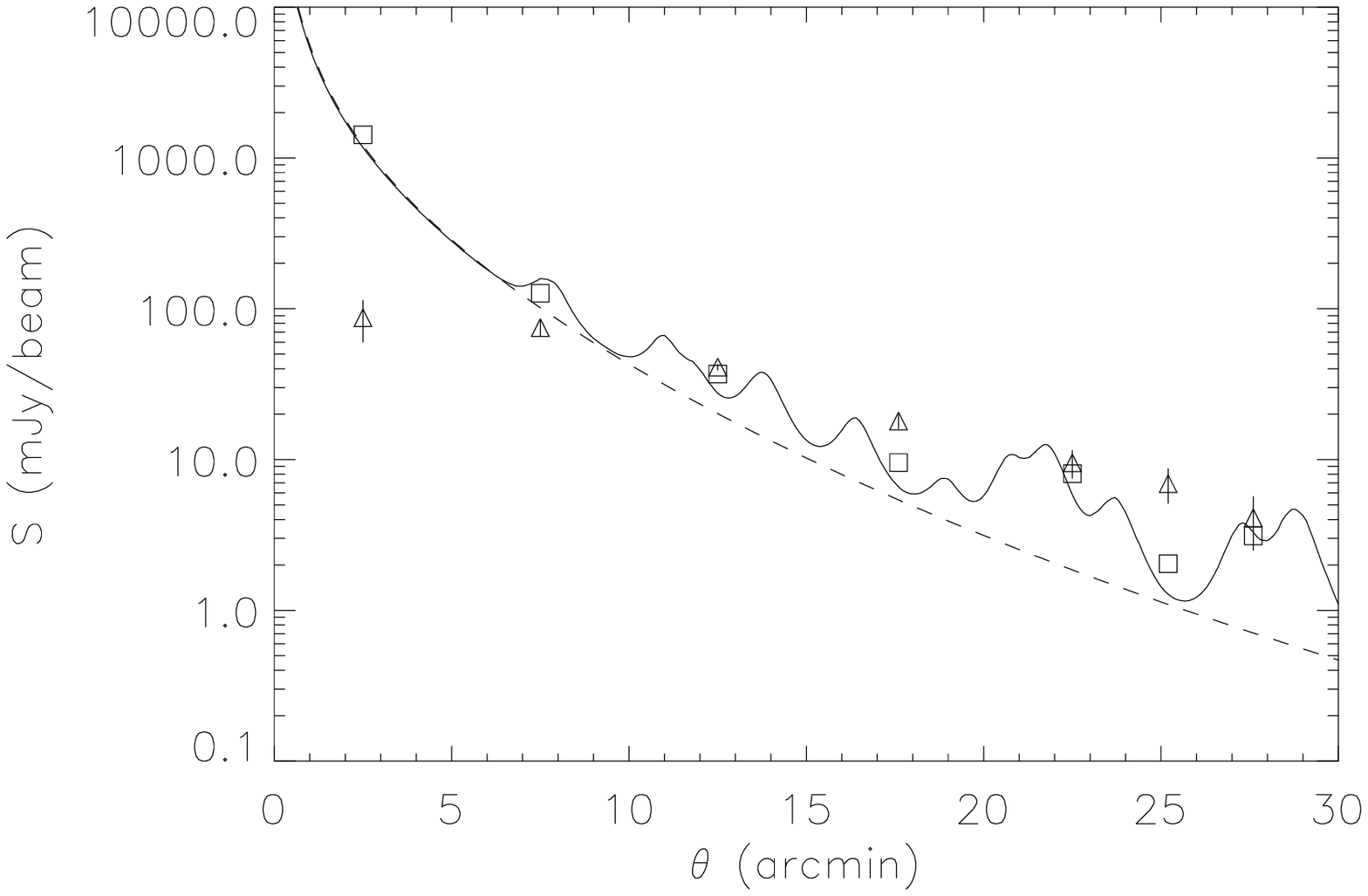,scale=0.6,angle=0.0}
}
\end{center}
 \caption{Azimuthally averaged radio surface brightness at 1.4 GHz for the same models than in Fig.\ref{radio_flux} (upper panel: 9 GeV model; lower panel: 43 GeV model). The solid line is the total emission, and the dashed line is the emission of the main halo. Triangles are data from Deiss et al. \cite{Deissetal1997}, squares are the average of the surface brightness calculated in concentric rings with the same centers and extensions than in Deiss et al. \cite{Deissetal1997}.}
 \label{radio_sb}
\end{figure}

We also calculated the azimuthally averaged radio surface brightness at 1.4 GHz, and we compared our result with the observed data of Deiss et al. \cite{Deissetal1997}. In Fig.\ref{radio_sb} the emission produced by all the DM halos is compared with the emission of the main halo and with the data plotted in Fig.3 of Deiss et al. \cite{Deissetal1997}. The surface brightness emissions for both the 9 and 43 GeV neutralino models look quite similar. In the same figure we also show with the squares the average of the radio surface brightness calculated in concentric rings having the same centers and extensions than in Deiss et al. \cite{Deissetal1997}. We notice that, outside the most internal circle within a radius of $\sim 5$ arcmin, the effect of the DM sub-halos is to produce an average surface brightness profile wider than the emission of the main DM halo alone, allowing to have a better agreement with the observed data out to large radii, that instead is not possible to obtain by considering the main DM halo alone (see discussion in \cite{Colafrancescoetal2011}). We also note that in the central part of the halo the removal of the radio emission from the cD galaxy NGC 4874 (that has a radius of $\sim2-3$ arcmin) could have altered the profile of the diffuse emission.

\subsection{Flux and surface brightness in X-rays}

Finally, we calculate here also the emission in the X-ray band that the non-thermal electrons produced in DM halos produce by inverse Compton scattering of the CMB photons and by non-thermal bremsstrahlung with the hot gas nuclei.

\begin{figure}
\begin{center}
{
 \epsfig{file=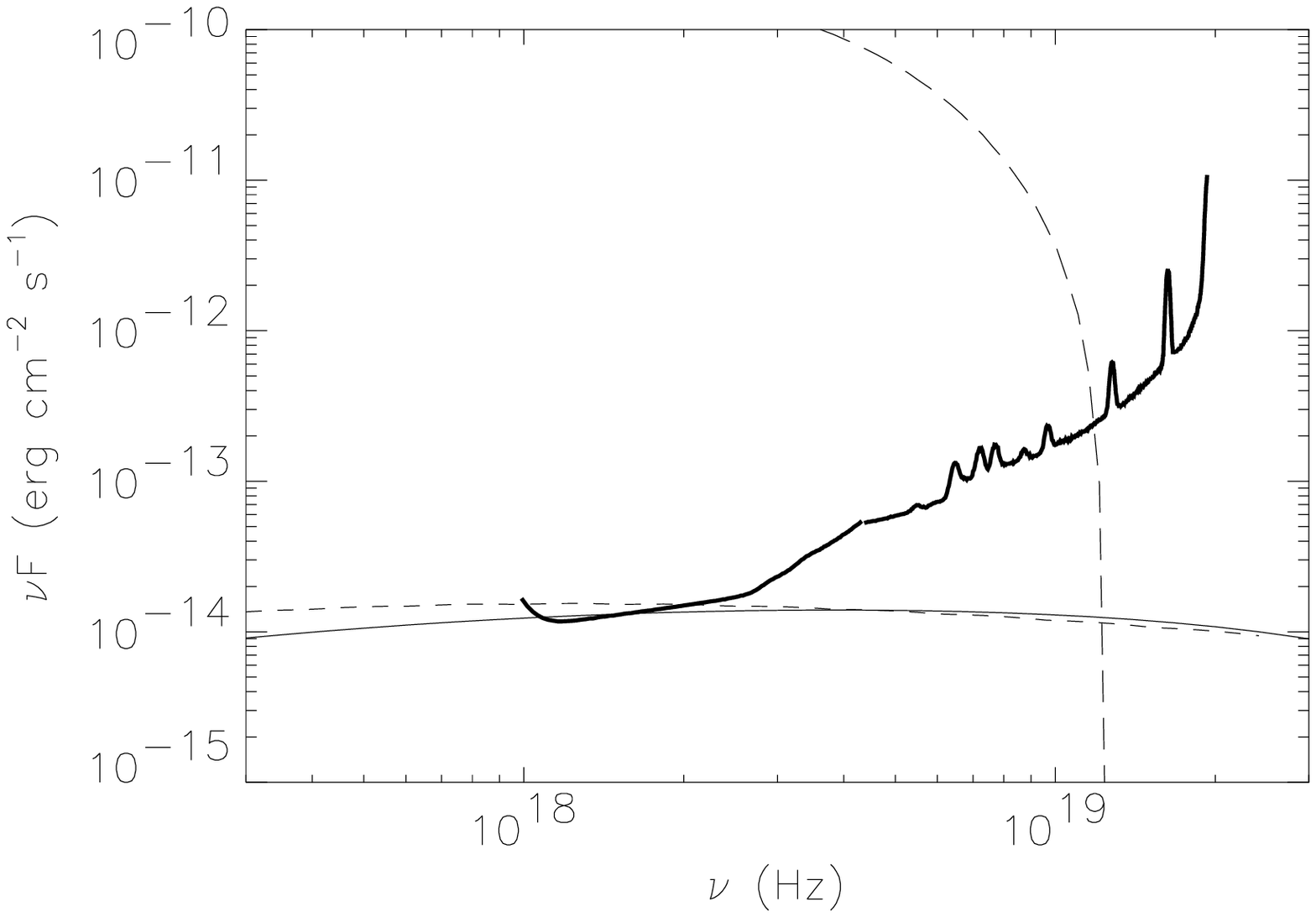,scale=0.6,angle=0.0}
}
\end{center}
 \caption{X-ray flux produced in Coma from the DM halos for a neutralino with $M_\chi=9$ GeV (solid line) and for $M_\chi=43$ GeV (dashed line) with the same properties than in Fig.\ref{radio_flux}. Long-dashed line is the thermal emission of the hot gas in Coma. The expected sensitivity of Astro-H HXI for 100 ks of time integration (from http://astro-h.isas.jaxa.jp/researchers/sim/sensitivity.html) is also plotted.}
 \label{xrays_flux}
\end{figure}

Figure \ref{xrays_flux} shows that an instrument like Astro-H HXI can in principle detect this emission at energies around 5 keV, but at this energy the non-thermal emission is completely dominated by the thermal one. The non-thermal emission becomes comparable to the thermal one at energies around 30 keV, but at this energy the expected flux is lower than the sensitivity of Astro-H by a factor 20 or more.

\begin{figure}
\begin{center}
{
 \epsfig{file=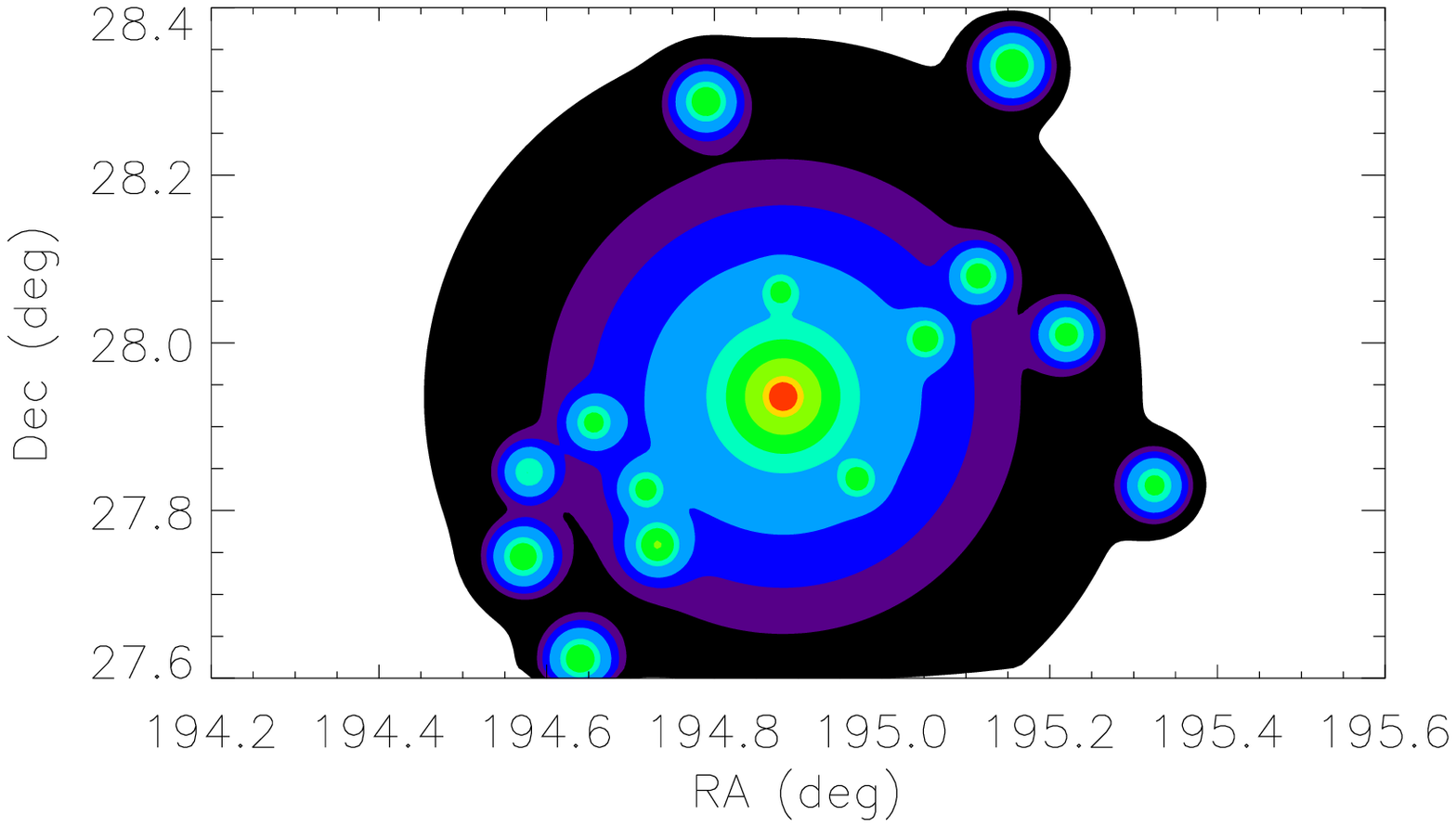,scale=0.7,angle=0.0}
}
\end{center}
 \caption{Map of the X-rays (integrated in the band 1--10 keV) surface brightness in Coma from the DM halos for the 9 GeV model in Fig.\ref{radio_flux}, smoothed on a scale of 1 arcmin.}
 \label{xrays_map}
\end{figure}

In Figure \ref{xrays_map} we show also the X-ray map we expect (smoothed on a scale of 1 arcmin) for the DM model with neutralino mass of 9 GeV (the case with 43 GeV is very similar). This map is what we expect once the thermal emission is removed; however this last emission is of the order of $10^4-10^5$ higher than the DM-induced  emission, and therefore the possibility to observe the DM emission is strongly reduced.

\section{Discussion and conclusions}

We have shown in this paper that the detailed inclusion of the observed DM sub-halos properties in the calculation of the non-thermal emission from the Coma cluster allows to reproduce the observed radio halo properties (both the spectrum and the surface brightness distribution) of Coma and is also consistent with the Fermi-LAT upper limits for this cluster, under assumptions consistent with the available information about the magnetic field properties obtained from Faraday Rotation measures and with the DM properties derived from observations of the Galactic center.

We showed in particular that the inclusion of DM sub-halos has two main effects:\\
1) to increase (even by a factor of $10-20\%$) the total flux of the DM-induced emission, allowing to reduce the gap between the DM annihilation cross section values necessary to reproduce the observed radio emission and the present upper limits derived from other observations; this gap can be recovered when optimistic (but still realistic) assumptions are used for the value of the substructure boost factor and, possibly, for the effect of the fluctuations in the magnetic field: in fact, it has been found that the inclusion of smaller sub-halos can boost the DM-induced signals by a factor that under optimistic assumptions can be of the order of $\sim100$, and that the fluctuations in the magnetic field structure can increase the radio signal by another factor $\sim1.5-2$;\\ 
2) to change the radio surface brightness distribution, allowing to obtain results more similar to the observed ones with respect to the case when only the effect of the DM main halo is considered. We have found that, while the resulting azimuthally averaged surface brightness profile is quite similar to the observed one, some difference is present when the full structure of the radio map is considered; however, we note that the model used to describe the DM sub-halo distribution, where the sub-halos are described as 15 independent spherically symmetric halos, is an approximation of the complex structure of the DM distribution \cite{Okabe2014}, and that the procedure to obtain the radio map, particularly regarding the subtraction of discrete and slightly extended sources, is quite delicate, therefore the difference between the observed and calculated maps can be due to these effects.

The results obtained in this paper are also interesting because DM models have other appealing features. First, the possibility to reproduce the steepening of the radio halo spectrum in the Coma cluster using the observed magnetic field properties; this is particularly true at high frequencies for the neutralino model with mass of 9 GeV and composition $\tau^+ \tau^-$. This model appears to be the most interesting one between the two considered in this paper, because it also allows to reproduce the intensity of the radio flux for reasonable values of the boost factor due to a realistic distribution of DM substructures, even though also the model with mass of 43 GeV and composition $b \bar b$ cannot be excluded. We note that these models are corresponding to the best fit cases found by Abazajian \& Keeley \cite{Abaza2016} from the Fermi data of the Galactic center, and are close to the model suggested by the data on the cosmic ray antiproton spectrum obtained with PAMELA, with mass of 35 GeV and composition $b \bar b$ \cite{Hooper2015}, whereas they are in tension with results from positron excess measured with AMS which instead point to mass of the order of 500 GeV or higher (e.g. \cite{Accardo2014}), that cannot fit the radio halo spectrum of Coma for reasonable values of the magnetic field and without violating the Fermi-LAT upper limits \cite{Colafrancescoetal2011,Beck2015}.

Another very interesting feature of DM models is the fact that they do not predict a level of acceleration of cosmic ray protons in galaxy clusters that produce a gamma-ray emission in excess with respect to the Fermi-LAT upper limits, unlike most of the other models presently proposed for the origin of radio halos in galaxy clusters. In fact, as pointed out by Vazza et al. \cite{Vazza2016}, the recent Fermi-LAT upper limits indicate a content of cosmic ray protons in galaxy clusters lower than the value previously derived from simulations of galaxy cluster merging (see, e.g., \cite{Ryu03}) and this result, in turn, strongly challenges our present understanding of the efficiency in the acceleration of cosmic rays produced by processes related to clusters mergers (like, e.g.,  shocks or turbulences), requiring therefore to find other sources of cosmic ray acceleration that can accelerate electrons without accelerating protons. DM models are perfectly fitting this requirement, and we have verified in this paper, for the DM models we have considered, and for the DM annihilation cross section values and boost factors necessary to reproduce the radio flux, that the gamma-ray upper limits are not violated. 

Therefore, we conclude that DM models are a viable possibility to explain the origin of radio halos in galaxy clusters, allowing to reproduce the spectral and spatial properties of the radio halos without producing a gamma-ray flux in excess of the Fermi-LAT limits, and without requiring new and more sophisticated (and still not well studied) models to understand the physics of acceleration of cosmic rays in galaxy clusters (see, e.g., \cite{Vazza2016} and references therein). Also, DM models do not require a strong fine tuning because, once the mass and composition of the neutralino are fixed, the only other free parameter in these models is given by the product of the annihilation cross section by the substructures boost factor; actually, all these parameters are also not completely free because they can be constrained from other observations or theoretical studies, like numerical simulations and the relative determination of the boosting factor.
  
Further analyses (both theoretical and observational) will be useful to clarify this possibility. On the theoretical side, more improved models derived from gravitational lensing and/or from numerical simulations, where the full spatial distribution of DM is considered, deserve to be better studied. Also the study of the DM emission in other clusters, and the study of the correlation between DM and radio properties in a number of cluster will be necessary. In particular, the presence of clusters with similar mass and/or dynamical state but very different radio powers, like e.g. the Bullet cluster and A2146 \cite{King2016}, is still an open issue in our understanding of the origin of cluster radio halos, and it is necessary to understand if this difference can be explained in the DM models, possibly by considering the properties and the structure of the DM sub-halos.

On the observational side, the independent experiments (both laboratory and astrophysical) devoted to constrain the nature and the properties of DM are important to understand the role of DM in the production of non-thermal emission in galaxy clusters. Regarding the possibility to obtain this information by direct observations of galaxy clusters, we point out that the best spectral bands to test these models are the radio and the gamma-ray bands.\\ 
In the radio band it is desirable to better constrain the spectral and the spatial properties of the diffuse radio emission in galaxy clusters through detailed observations with high angular resolution and high sensitivity of the surface brightness distribution at different frequencies, together with measures of the intra cluster magnetic field; this appears to be the best and most feasible possibility to constrain the DM models, because these observations can be performed with a forthcoming instrument like the SKA.\\ 
In order to test these models in the gamma-ray band,  it would be necessary to see if, with the data coming from the full operative lifetime of Fermi and the new techniques of data analysis (like the Pass-8), it is possible to improve the present Fermi-LAT upper limits by a factor of the order of 2--3 at energies around 1 GeV. We note also that Cerenkov telescopes operating in the very high-energy band ($E>100$ GeV) are not expected to detect the DM emission for the values of the neutralino mass we considered in this paper; the DM emission in this band can be eventually observed only for high values of the neutralino mass (e.g., $M_\chi>1$ TeV), but in this case there is not the possibility to reproduce the shape of the radio spectrum (see, e.g., \cite{Beck2015}).\\
The possibility to detect the DM predicted emissions in the X-ray band is harder, because these emissions are slightly higher than the sensitivity limits of the next generation telescopes only at energies where the non-thermal emission is dominated by the thermal one. In the hard X-ray and soft gamma-ray bands the emission coming from DM-produced electrons is well below the sensitivity of the next generation and/or proposed instruments like Astro-H2 or AstroMeV and ASTROGAM (Colafrancesco \& Marchegiani, in preparation): however, a possible detection of galaxy clusters in these bands can be used to rule out DM models.\\ 
The same is true for the microwave band, where DM models do not provide a strong non-thermal Sunyaev-Zel'dovich effect (SZE) signal because they produce a very low amount of electrons at low energies (see, e.g., \cite{Marchegiani2015}); therefore, a possible detection of the non-thermal SZE in Coma would allow to exclude a DM origin for the non-thermal electrons.

\section*{Acknowledgments}

This work is based on the research supported by the South African Research Chairs Initiative of the Department of Science and Technology and National Research Foundation of South Africa (Grant No 77948).
P.M. acknowledges support from the DST/NRF SKA post-graduate bursary initiative.
We thank the Referee for useful comments and suggestions.\\
\textit{Disclaimer: any opinion, finding and conclusion or recommendation expressed in this material is that of the author(s) and the NRF does not accept any liability in this regard}.

\end{document}